# CHANNEL ADVERSARIAL TRAINING FOR CROSS-CHANNEL TEXT-INDEPENDENT SPEAKER RECOGNITION

*Xin Fang*[1, 2], *Liang Zou*[3], *Jin Li*[2], *Lei Sun*[1] and *Zhen-Hua Ling*[1*]

University of Science and Technology of China, Hefei, P. R. China[1]
iFlytek Research, Hefei, P. R. China[2]
School of Electrical and Information Engineering, Anhui University, Hefei, P. R. China[3]

## ABSTRACT

The conventional speaker recognition frameworks (e.g., the i-vector and CNN-based approach) have been successfully applied to various tasks when the channel of the enrolment dataset is similar to that of the test dataset. However, in real-world applications, mismatch always exists between these two datasets, which may severely deteriorate the recognition performance. Previously, a few channel compensation algorithms have been proposed, such as Linear Discriminant Analysis (LDA) and Probabilistic LDA. However, these methods always require the collections of different channels from a specific speaker, which is unrealistic to be satisfied in real scenarios. Inspired by domain adaptation, we propose a novel deep-learning based speaker recognition framework to learn the channel-invariant and speaker-discriminative speech representations via channel adversarial training. Specifically, we first employ a gradient reversal layer to remove variations across different channels. Then, the compressed information is projected into the same subspace by adversarial training. Experiments on test datasets with 54,133 speakers demonstrate that the proposed method is not only effective at alleviating the channel mismatch problem, but also outperforms state-of-the-art speaker recognition methods. Compared with the i-vector-based method and the CNN-based method, our proposed method achieves significant relative improvement of 44.7% and 22.6% respectively in terms of the Top1 recall.

***Index Terms***— *cross channel, speaker recognition, channel adversarial training*

## 1. INTRODUCTION

With the popularity of smartphones and mobile devices, speaker recognition has attracted more and more attentions, as its non-contact, low-cost and other advantages. Given the fact that personal speech is always stored and transmitted on different hardware and applications, a crucial issue need to address is cross-channel speaker recognition.

Conventional frameworks of speaker recognition, such as i-vector-based strategies, have been successful in the last decade [1, 2]. They always assume that the enrolment and test dataset share the same distribution. Unfortunately, this assumption does not hold in many real-world applications because there is often channel mismatch between enrolment and test data. The channel mismatch significantly affects the speaker recognition performance. To address the mismatch challenge, several techniques have been developed and achieved state-of-the-art performance such as Linear Discriminant Analysis (LDA) [1], Probabilistic LDA (PLDA) [3] and so on. Channel compensation is regarded as a potential solution to mitigate the mismatch problem. It has attracted great interest of researchers in the field of speaker recognition due to its remarkable performance.

The channel compensation methods for systems using i-vector back-ends have been the dominating paradigm of channel compensation. Recently, deep learning is becoming a mainstream technology for speech recognition [4]. Many efforts have been made using deep neural networks (DNNs) to compensate the channel mismatch for speaker recognition. Currently, the most promising approaches are end-to-end embedding architectures such as the deep speaker [5]. It has shown that Convolutional Neural Networks (CNNs) can achieve better performance than DNNs for integrated end-to-end architectures in text-independent speaker recognition scenarios [6, 7]. However, for a single speaker, it is difficult to collect training data from different channels. Therefore, the models are difficult to represent speaker information between different channels, which is one unsolved challenge in traditional channel compensation techniques. Inspired by unsupervised domain adaptation [8, 9, 10], we propose to learn the channel-invariant and speaker-discriminative speech representations via channel adversarial training (*CAT*) which only needs the labeled data under their respective ch annels. Moreover, unlike the unsupervised domain adaptive speaker recognition in the i-vector space [9], we directly conduct channel adversarial training under the CNN-based speaker recognition framework to solve the problem of channel mismatch. We further compares the performance of our proposed *CAT* method with state-of-the-art channel compensation methods. Experimental results on a large datasets with 54,133 speakers demonstrate that the proposed *CAT* method achieves the best performance.

[*] Zhen-Hua Ling is the corresponding author.

## 2. MOTIVITION

In the field of speaker recognition, the channel variability (i.e., mismatch between enrolment and test datasets) is one of the enduring challenges and is a major cause of errors. The variability arises from intrinsic factors (e.g., speaker characteristics) and extrinsic factors (e.g., how the speech is collected). For instance, the speech recorded by the software *A* is used for speaker enrolment, and the speech recorded by the software *B* is used for the speaker recognition. Generally, the speech codecs, also referred as channels, between different software tend to have large differences. The mismatch across channels can significantly degrade the speaker recognition performance. Typical channel compensation algorithms require a large amount of speech data to capture the information of the same speaker under different channels, especially when the inter-channel variability is large. However, it is unrealistic to collect cross channel data from a specific speaker in real and practical scenarios, which limits the utility of speaker recognition in many applications. Inspired by the work in domain adaptation, we design a robust speaker recognition framework which can well address the channel mismatch by adversarial training. By integrating this method, we expect that the speaker recognition technology can be practically applied in far more scenarios where we do not need to consider the difference across channels.

## 3. BASELINE ARCHITECTURE

### 3.1 I-vector and Modified Channel Compensation Methods

The i-vector based framework was originally proposed by Dehak et al. [1] and has recently become a popular strategy for text-independent speaker recognition [11]. It assumes that the speaker information can be modeled via the Gaussian Mixture Model (GMM) vectors. An efficient way to estimate the total-variability subspace and the subsequent i-vector is described by Kenny et al. [12] and Dehak et al. [13]. More recently, a few standard channel compensation techniques have been explored, such as LDA and probabilistic PLDA to model the channel variability within the i-vector space [11,14].

### 3.2 Basic Convolutional Architecture

The structure of the baseline CNN model includes five convolutional layers, which is same as the D1 module in Fig.1. For the input layer, 500 frames of 64-dimensional filter-bank features, which belong to the same person are grouped together as a feature map. Kernel size of each convolutional layer is 3x3, and the stride is set to be 1. Each convolutional layer is connected to pooling layer of 2x2 max pooling. Finally, the average pooling is used to get the speaker representation embedding, and the softmax loss as well as the triplet loss are employed for training.

#### *3.2.1 Loss Function*

The total loss is a combination of the softmax loss and the triplet loss [5]. The softmax loss is defined as:

$$L_s = -\sum_{i=1}^{M} \log \frac{e^{W_{y_i}^T x^i + b_{y_i}}}{\sum_{j=1}^{N} e^{W_j^T x^i + b_j}} . \quad (1)$$

where $x^i$ denotes the *i*-th speaker embedding, belonging to the $y_i$ speaker. $W_j$ denotes the *j*-th column of the weights matrix $W$ in the last fully connected layer and $b$ is the bias term. The size of mini-batch and the number of speakers is $M$ and $N$, respectively. The triplet loss is defined as:

$$L_T = \sum_{i=1}^{M} \max(0, D(x^i, x^n) + \delta - D(x^i, x^p)) . \quad (2)$$

The triplet loss is calculated via triplet of training samples $(x^i, x^n, x^p)$, where $(x^i, x^p)$ have the same speaker labels and $(x^i, x^n)$ have different speaker labels. $x^i$ is usually taken as an anchor of the triplet. Intuitively, the triplet loss encourages the model to find an embedding space where the distances between samples from the same speaker are smaller than those from different speaker by at least a margin $\delta$. $D(*,*)$ represents the cosine distance between two input vectors. Finally, the softmax loss and the triplet loss are combined together with a weight $\alpha$ to construct the total loss, shown as,

$$L = L_s + \alpha L_T. \quad (3)$$

#### *3.2.2 Recognition*

Given the trained network, the utterance-level embedding of the enrolment and test utterances are extracted from the basic CNN model. Specifically, if the duration of an utterance is shorter than the duration of the input segments utilized at the training stage, we pad some frames to the short utterance. Otherwise, we divide the long utterance into multiple short segments by employing a sliding window without overlap. Then the utterance-level speaker embedding is obtained by performing averaging pooling followed by *L2* normalization. After extracting the utterance-level speaker embedding, cosine distance is adopted as the scoring method.

## 4. CHANNEL ADVERSARIAL TRAINING

We propose to project two different channels into a common subspace to eliminate the channel mismatch. This can be achieved by training a model that learns a speaker-discriminative and channel-invariant feature representation. Inspired by the basic convolutional network described in Section 3.2, we extend that idea and propose a novel *CAT* architecture, as shown in Figure 1. The *CAT* is different from the CNN model in two folds. First, we add a generator network which has two LSTM layers. Second, we add a discriminator to predict the channel label, denoted as $d_i$ ([0,1] or [1,0]) for the *i*-th sample, which indicates whether $x^i$ comes from channel *A* or channel *B*. This model can be decomposed into three parts to perform different mappings, including a feature extractor *G*, a speaker label predictor *D1* and a channel predictor *D2*. More formally, the mapping functions can be expressed as

$$G = f_G(x, \theta_G), \tag{4}$$

$$D1 = f_{D1}(g, \theta_{D1}), \tag{5}$$

$$D2 = f_{D2}(g, \theta_{D2}), \tag{6}$$

where $\theta_G$ , $\theta_{D1}$, $\theta_{D2}$ are the parameters of the network (in Figure 1) . Our aim is to jointly train $\theta_G$, $\theta_{D1}$ and $\theta_{D2}$. Specifically, we want to optimize $\theta_G$ by minimizing the speaker label prediction loss and maximizing the channel classification loss at the same time, which can be realized by a gradient reversal layer. Gradient reversal layer between the feature extractor and channel label predictor is introduced to search the saddle point between speaker label classifier and channel classifier. We multiply the gradients with $\beta$ during the backpropagation, as shown in Eq. (11). $\beta$ is a positive hyper parameter used to trade off the *D1* loss and *D2* loss in practice. Gradient reversal layer ensures the feature distributions over the two channels are similar so that we can get channel-invariant and speaker-discriminative features.

### 4.1. Loss Function

The total loss is a combination of the losses from *D1* and *D2*. The *D1* loss is defined as Eq. (3) in section 3.2.1,

$$L_{D1} = L_s + \alpha L_T. \tag{7}$$

The *D2* loss is defined as,

$$L_{D2} = -\sum_{i=1}^{M} \log \frac{e^{W_{d_i}^T x^i + b_{d_i}}}{\sum_{j=1}^{K} e^{W_j^T x^i + b_j}}, \tag{8}$$

where $x^i$ denotes the *i*-th speaker embedding, belonging to the $y_i$ speaker. $W_j$ denotes the *j*-th column of the weights *W* in the last fully connected layer and *b* is the bias term. The size of minibatch and the number of channel is *M* and *K*. The overall *CAT* network is optimized via stochastic gradient descent (SGD) [15,16] approach. The optimal parameters are achieved through the following two equations,

$$(\hat{\theta}_G, \hat{\theta}_{D1}) = \underset{\theta_G, \theta_{D1}}{\arg\min} E(\theta_G, \theta_{D1}, \hat{\theta}_{D2}), \tag{9}$$

$$(\hat{\theta}_{D2}) = \underset{\theta_{D2}}{\arg\max} E(\hat{\theta}_G, \hat{\theta}_{D1}, \theta_{D2}). \tag{10}$$

The optimization formulas can be written as

$$\theta_G = \theta_G - l * \left(\frac{\partial L_{D1}}{\partial \theta_G} - \beta * \frac{\partial L_{D2}}{\partial \theta_G}\right), \tag{11}$$

$$\theta_{D1} = \theta_{D1} - l * \left(\frac{\partial L_{D1}}{\partial \theta_{D1}}\right), \tag{12}$$

$$\theta_{D2} = \theta_{D2} - l * \left(\frac{\partial L_{D2}}{\partial \theta_{D2}}\right). \tag{13}$$

After model training, we can extract channel-invariant and speaker-discriminative features extracted from the neural network.

## 5. EXPERIMENTS

### 5.1. Speech Data

Experiments were performed on a large collection of speakers from four homemade sessions in iFlytek Co., Ltd.

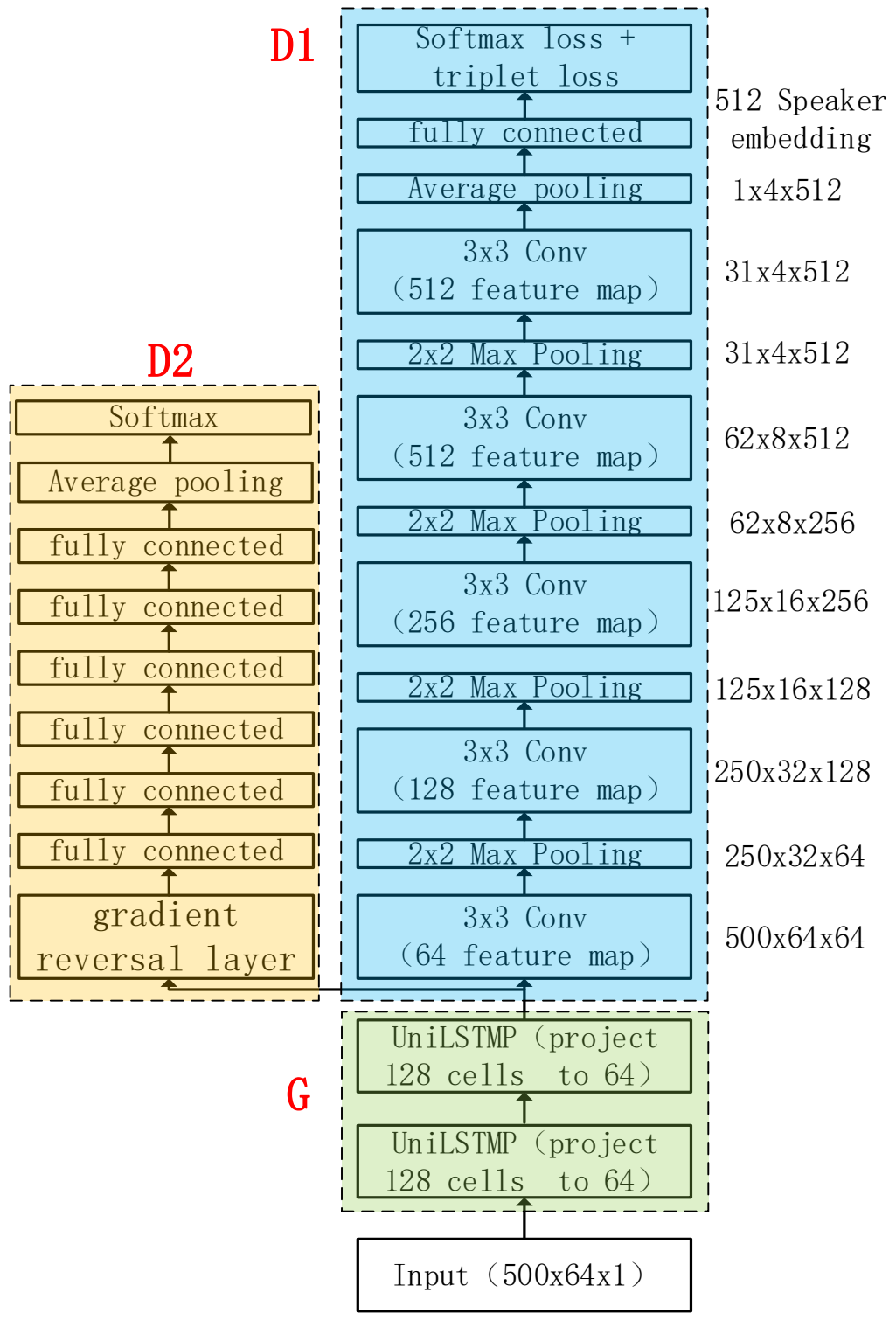

**Figure 1:** Architecture of channel adversarial training model.

**Training Set1**: The codec used in this dataset is Speex [17], which is a patent-free audio compression format designed for speech. Speex is a common way for speech codec. This set includes 37,557 speakers, each speaker has 60 utterances on average. Utterance duration is 8 seconds on average.
**Training Set2**: The codec used in this dataset is SILK [18], which is also a patent-free audio compression format. This set includes 38,046 speakers, each speaker has 30 utterances on average. Utterance duration is 12 seconds on average.
**Development Set**: All utterances from the other 22 speakers were used as validation set for adjusting the parameters. For each speaker, one utterance was sampled by SILK codec as the enrolment data. Besides, we collected the other 25 speech by Speex codec as the cross-channel test data. This resulted in 550 target trials and 11,550 impostor trials in total.
**Test Set**: All of the utterances from the other 54,133 speakers were used as the test set for evaluating the systems' performance. For each speaker, 10 utterances were sampled by SILK codec as the enrolment data. We further sampled 246 utterances by Speex codec from 100 speakers included in the enrolled 54,133 speakers as the cross-channel test data. This resulted in 246 target trials and 13,316,472 impostor trials in total.

### 5.2. Evaluation Metrics

Because the test set was too large and the number of test data was tens of millions, we performed a cross-channel speaker recognition task on a smaller development set to adjust the parameters. The Equal Error Rate (EER), which corresponds to the threshold where the probability of miss-classifying positive samples is same as that for negative samples, was used to evaluate the performance in the development set. We further verified the performance of speaker recognition task on a large-scale test data. The performance index used for the test set was TopN recall rate. Supposing we need to judge which speaker the given speech belong to among the $S$ speakers, the test speech is compared with the $S$ speakers. If the targeted speaker exists among the most similar $N$ speakers, we can consider it as a successful recall. The TopN recall rate can be calculated as follows,

$$\text{TopN recall} = \frac{\text{the number of successful recall speeches}}{\textit{the number of test speeches}}. \quad (14)$$

### 5.3. Model training

The i-vector extractor was trained by training set1 and training set2. It was based on a UBM with 512 Gaussian mixtures and a gender-independent total variability matrix with 300 total factors. We employed within-class covariance normalization (WCCN) [19] and i-vector length normalization (LN) [20] to the 300-dimensional i-vector. Then the LDA and WCCN were used to further alleviate intra-speaker variability and reduce the dimension to 200. Finally, PLDA models with 150 latent identity factors were trained.

The basic convolutional neural network and the proposed channel adversarial training method were trained by training set1 and training set2. We employed the SGD [16, 17] optimizer with an initial learning rate of 0.2 for all network components. The learning rate was decayed based on the performance on the development set. To accelerating the training process, batch normalization and dropout were employed during the training process. The batch size was set to 64 and the value of $\alpha$ in Eq. (3) was set to 1.

### 5.4. The effect of $\beta$ in *CAT*

We investigated the impact of the hyper-parameters $\beta$, which was used to achieve a tradeoff between two sub-losses, on the performance of the proposed *CAT* method. The impact of $\beta$ on EER on development set and Top1 recall on test set are depicted in Fig. 2. The lowest EER and the highest Top1 recall were achieved when $\beta$ was set to be 1.

### 5.5 Performance comparison between different methods

The experimental results are shown in Table 1. The *CAT without D2* represents the channel adversarial training system without channel classification. The purpose of *CAT without D2* is to ensure that its model complexity is completely comparable to system *CAT* without using the channel adversarial training method.

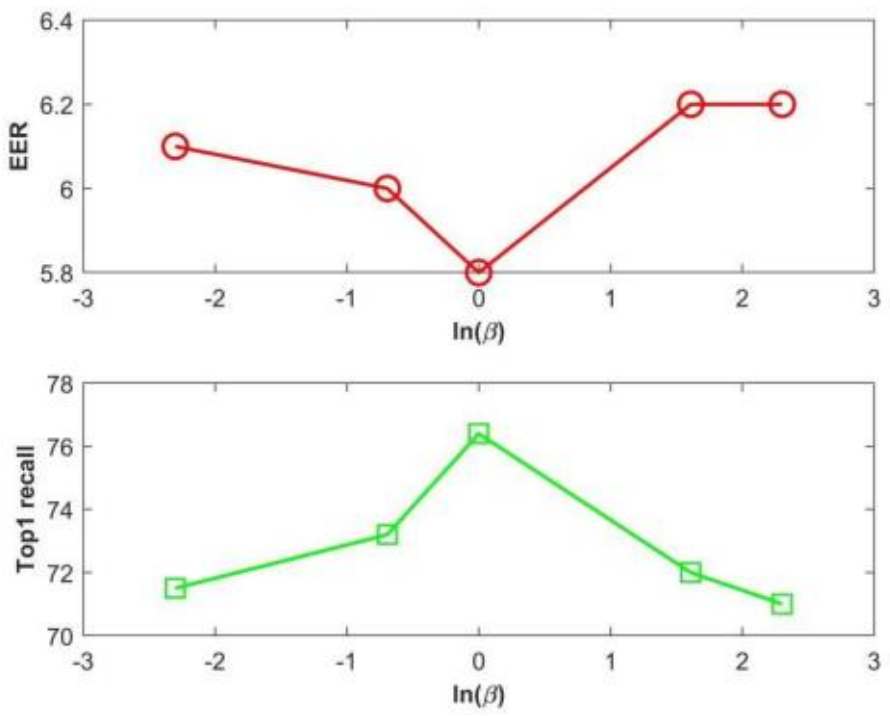

**Figure 2:** The EER and Top1 recall of *CAT* as a function of $\beta$

**Table 1. The EER (%) of *CAT* and state-of-the-art methods on the development set.**

| System | i-vector | CNN | *CAT without D2* | *CAT* |
|---|---|---|---|---|
| EER(%) | 8.8 | 6.2 | 6.4 | 5.8 |

Comparing CNN with *CAT without D2*, it can be seen from Table 1 that only adding the feature extraction generator G cannot improve the performance. This further demonstrates the effectiveness of the proposed method.

**Table 2. The TopN recall rates (%) of *CAT* and state-of-the-art methods on the test set.**

| System | Top1 | Top5 | Top10 |
|---|---|---|---|
| i-vector | 57.3 | 66.3 | 70.3 |
| CNN | 69.5 | 77.6 | 80.1 |
| *CAT without D2* | 69.1 | 78.0 | 80.1 |
| *CAT* | 76.4 | 83.3 | 85.0 |

As can be seen from Table 2, by projecting the data to a common space using the proposed *CAT* approach, we achieved 22.6% relative improvement (absolute improvement 6.9%) over the CNN baseline system on Top1 recall. The result on the test set was consistent with that on the development dataset, which further shows the robustness of the proposed *CAT* strategy.

## 6. CONCLUSIONS

In this paper, we propose a cross-channel speaker recognition approach based on channel adversarial training, which alleviates the channel mismatch problem by projecting the data of two channels into the same subspace. Through this strategy, we can obtain channel-invariant and speaker-discriminative speech representations. Experiments on a large test dataset show that, the proposed approach improves the Top1 recall rate from 69.5% to 76.4%, with 22.6% relative improvement. In the future, we will explore the *CAT* method on the datasets with more than two channels.

## REFERENCES

[1] N. Dehak, P. J. Kenny, R. Dehak, *et al.*, "Front-End Factor Analysis for Speaker Verification," *Ieee Transactions on Audio Speech and Language Processing,* vol. 19, pp. 788-798, May 2011.

[2] H. Zeinali, H. Sameti, L. Burget, *et al.*, "Text-dependent speaker verification based on i-vectors, Neural Networks and Hidden Markov Models," *Computer Speech and Language,* vol. 46, pp. 53-71, Nov 2017.

[3] S. J. D. Prince and J. H. Elder, "Probabilistic linear discriminant analysis for inferences about identity," *2007 Ieee 11th International Conference on Computer Vision, Vols 1-6,* pp. 1751-1758, 2007.

[4] Y. Bengio, "Learning deep architectures for AI," *Foundations and trends® in Machine Learning,* vol. 2, pp. 1-127, 2009.

[5] C. Li, X. Ma, B. Jiang, *et al.*, "Deep speaker: an end-to-end neural speaker embedding system," *arXiv preprint arXiv:1705.02304,* 2017.

[6] S. Wang, Y. M. Qian and K. Yu, "Focal Kl-Divergence Based Dilated Convolutional Neural Networks for Co-Channel Speaker Identification," *2018 Ieee International Conference on Acoustics, Speech and Signal Processing (ICASSP),* pp. 5339-5343, 2018.

[7] S. Shon, H. Tang and J. Glass, "Frame-level speaker embeddings for text-independent speaker recognition and analysis of end-to-end model," *arXiv preprint arXiv:1809.04437,* 2018.

[8] Y. Ganin, E. Ustinova, H. Ajakan, *et al.*, "Domain-Adversarial Training of Neural Networks," *Journal of Machine Learning Research,* vol. 17, 2016.

[9] Q. Wang, W. Rao, S. N. Sun, *et al.*, "Unsupervised Domain Adaptation Via Domain Adversarial Training for Speaker Recognition," *2018 IEEE International Conference on Acoustics, Speech and Signal Processing (ICASSP),* pp. 4889-4893, 2018.

[10] L. Zhao, Z. Chen, Y. Yang, *et al.*, "ICFS Clustering With Multiple Representatives for Large Data," *IEEE transactions on neural networks and learning systems,* pp. 1-11, 2018.

[11] A. Kanagasundaram, D. Dean, S. Sridharan, *et al.*, "I-vector based speaker recognition using advanced channel compensation techniques," *Computer Speech and Language,* vol. 28, pp. 121-140, Jan 2014.

[12] P. Kenny, P. Ouellet, N. Dehak, *et al.*, "A study of interspeaker variability in speaker verification," *IEEE Transactions on Audio, Speech, and Language Processing,* vol. 16, pp. 980-988, 2008.

[13] N. Dehak, R. Dehak, J. Glass, *et al.*, "Cosine Similarity Scoring without Score Normalization Techniques," *Odyssey 2010: The Speaker and Language Recognition Workshop,* pp. 71-75, 2010.

[14] P. Matejka, O. Glembek, F. Castaldo, *et al.*, "Full-Covariance Ubm and Heavy-Tailed Plda in I-Vector Speaker Verification," *2011 IEEE International Conference on Acoustics, Speech, and Signal Processing,* pp. 4828-4831, 2011.

[15] Y. LeCun, Y. Bengio and G. Hinton, "Deep learning," *nature,* vol. 521, p. 436, 2015.

[16] L. Bottou, "Large-scale machine learning with stochastic gradient descent," in *Proceedings of COMPSTAT'2010*, ed: Springer, 2010, pp. 177-186.

[17] (2018). *Speex: A Free Codec For Free Speech*. Available: https://www.speex.org/

[18] (2018). *SILK*. Available: https://en.wikipedia.org/wiki/SILK

[19] A. O. Hatch, S. Kajarekar and A. Stolcke, "Within-Class Covariance Normalization for SVM-based Speaker Recognition," *Interspeech 2006 and 9th International Conference on Spoken Language Processing, Vols 1-5,* pp. 1471-1474, 2006.

[20] D. Garcia-Romero and C. Y. Espy-Wilson, "Analysis of I-vector Length Normalization in Speaker Recognition Systems," *12th Annual Conference of the International Speech Communication Association 2011 (Interspeech 2011), Vols 1-5,* pp. 256-259, 2011.